\documentclass[%
 reprint,
 amsmath,amssymb,
 aps,twocolumn
]{revtex4}

\usepackage{graphicx}
\usepackage{dcolumn}
\usepackage{bm}
\usepackage[utf8]{inputenc}
\usepackage[T1]{fontenc}
\usepackage{hyperref}
\usepackage{amssymb}
\usepackage{amsmath}
\usepackage[scr=rsfso,cal=zapfc,frak=euler,bb=ams]{mathalfa}
\usepackage{caption}
\usepackage{subcaption}
\usepackage{tikz}
\usepackage{xcolor}
\usepackage{soul}

\newcommand{\ket}[1]{\vert #1 \rangle}
\newcommand{\bra}[1]{\langle #1 \vert}
\newcommand{\ii}{\mathrm{i}}

\DeclareMathAlphabet\mathbfcal{OMS}{cmsy}{b}{n}

\begin{document}

\preprint{APS/123-QED}

\title{Generalized Cluster Correlation Expansion theory for STIRAP processes in the presence of a spin bath}

\author{Tommaso Fazio$^1$}%
 \email{tommaso.fazio@unipa.it}
\author{Anna Napoli$^{1,2}$}
\author{Benedetto Militello$^{1,2}$}
\affiliation{%
 $^1$Università degli Studi di Palermo, \\
 Dipartimento di Fisica e Chimica - Emilio Segrè, Via Archirafi 36, 90123 Palermo, Italia
}%
\affiliation{%
 $^2$I.N.F.N. Sezione di Catania, Via Santa Sofia 64, I-95123 Catania, Italia
}%


\date{\today}

\begin{abstract}
\noindent The Stimulated Raman Adiabatic Passage (STIRAP) is applied to a  system coupled to a bath made of fully-interacting two-level systems, whose dynamics is studied exploiting the generalized Cluster Correlation Expansion (gCCE) theory. We specialize our analysis to a negatively charged silicon vacancy ($\mathrm{V}_{\mathrm{Si}}^{-1}$) in non-purified 4H-SiC to assess the possibility of transferring population between two states of the ground manifold, also taking into account the interaction with a spherical nuclear spin bath formed by nuclei of $^{29}$Si and $^{13}$C. For this system, it is demonstrated that the presence of a small/medium sized bath has no effect on the protocol, finding in particular a set of parameter values for an efficient STIRAP process.
\end{abstract}

\maketitle


\section{\label{sec:intro}Introduction}

Stimulated Raman Adiabatic Passage (STIRAP) is a solid technique for quantum state manipulation of microscopic systems~\cite{vitanov2001coherent,vitanov2001laser,bergmann1998coherent,kral2007colloquium,bergmann2019roadmap}. It has a wide range of applications in several physical contexts, such as condensed matter~\cite{klein2007robust,alexander2008stimulated,golter2014optically,yale2016optical,wolfowicz2021quantum,zhou2017accelerated,baksic2016speeding} and plasmonic systems~\cite{varguet2016dressed,castellini2018quantum}, but also trapped ions~\cite{sorensen2006efficient,higgins2017coherent} and superconducting devices~\cite{kubo2016turn,kumar2016stimulated}. It consists in an adiabatic manipulation of a quantum system, usually effectively described as a three-state system, where the initial state and the target state (where we want to transfer population) are coupled to an auxiliary state through suitable pulses, realizing a Raman coupling with fields having time-dependent amplitudes. Typically, the most efficient way to realize a complete population transfer is through the so called counterintuitive sequence, where the coupling pulse between the target and the auxiliary state (Stokes pulse) precedes the coupling between the auxiliary and the initial state (pump pulse). Also the inverse sequence is possible, where the pump pulse precedes the Stokes one. This technique is addressed as bright STIRAP (b-STIRAP) and is more fragile to environmental effects. Because of the wide range of application of STIRAP, the effects of the environment have been extensively investigated, in order to understand at which extent this manipulation technique can be considered efficient in real experiments. Since the most common source of dissipation and decoherence is the interaction with the electromagnetic field, an extensive analysis of the action of a bosonic (photonic) environment is present in the literature, which has been developed through different mathematical treatments, ranging from an effective description of dissipation and decoherence via non-Hermitian Hamiltonians~\cite{vitanov1997population} to more appropriate treatments based on master equations~\cite{ivanov2005spontaneous,scala2010stimulated,scala2011stimulated,militello2011zeno,mathisen2018liouvillian}. Another important source of noise is the presence of fluctuations in the parameters characterizing the system, especially the pulses, whose effects on STIRAP manipulations have been studied as well~\cite{genov2013dynamical,yatsenko2014detrimental}.

\begin{figure}
\includegraphics[width=0.9\linewidth]{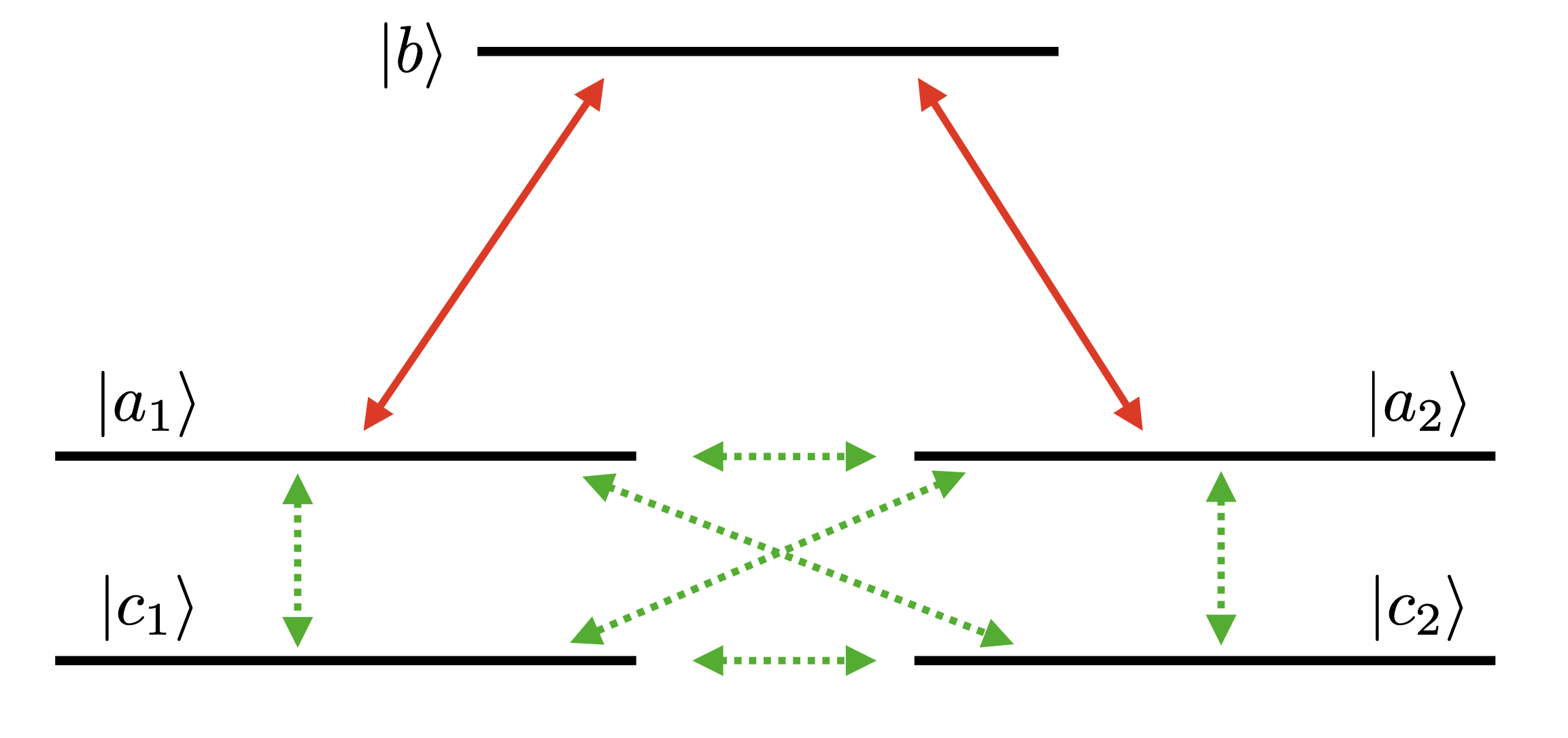}
\caption{Coupling scheme for the five-state system: $\ket{a_1}$ and $\ket{a_2}$ are the initial and target states, $\ket{b}$ is the auxiliary state involved in coherent coupling though the pump and Stokes pulses, represented by the solid red arrows. The four states $\ket{a_1}$, $\ket{a_2}$, $\ket{c_1}$ and $\ket{c_2}$ are in general connected through the interaction with the bath.}
\label{fig:generalscheme}
\end{figure}

Since STIRAP has applications in solid state physics, the effect of neighboring atoms is sometimes relevant, which justifies the inclusion in the model of an interacting bath of few-state systems. This is the case, for example, of diamond Nitrogen Vacancies~\cite{golter2014optically,zhou2017accelerated,baksic2016speeding} and rare-earth doped crystals~\cite{klein2007robust,alexander2008stimulated}. This motivates the study of STIRAP in the presence of an interaction of the system with a spin bath. In fact, recently, theoretical studies on the effects of a spin bath on STIRAP processes have been reported~\cite{militello2023adiabatic,militello2024adiabatically}. These contributions refer to specific situations allowing for particular approximations: the homogeneous model~\cite{militello2023adiabatic} and the weak coupling regime~\cite{militello2024adiabatically}. Moreover, both works refer to a coupling scheme where each manipulated state (the initial and the target) is coupled to the auxiliary state through the environment, but this is not the only possible system-environment scheme. In this paper we focus on a different scheme where each manipulated state is coupled to a state other than the auxiliary one (see Fig.~\ref{fig:generalscheme}), thus making the system an effective five-state system. This new scheme could possibly apply to different physical scenarios and surely fits the case of a particular defect in silicon carbide (SiC), as clarified in the following.

\begin{figure*}
\centering
\begin{subfigure}{0.4944\linewidth}
\caption{}
\label{subfig:defect}
    \includegraphics[width=0.6\linewidth]{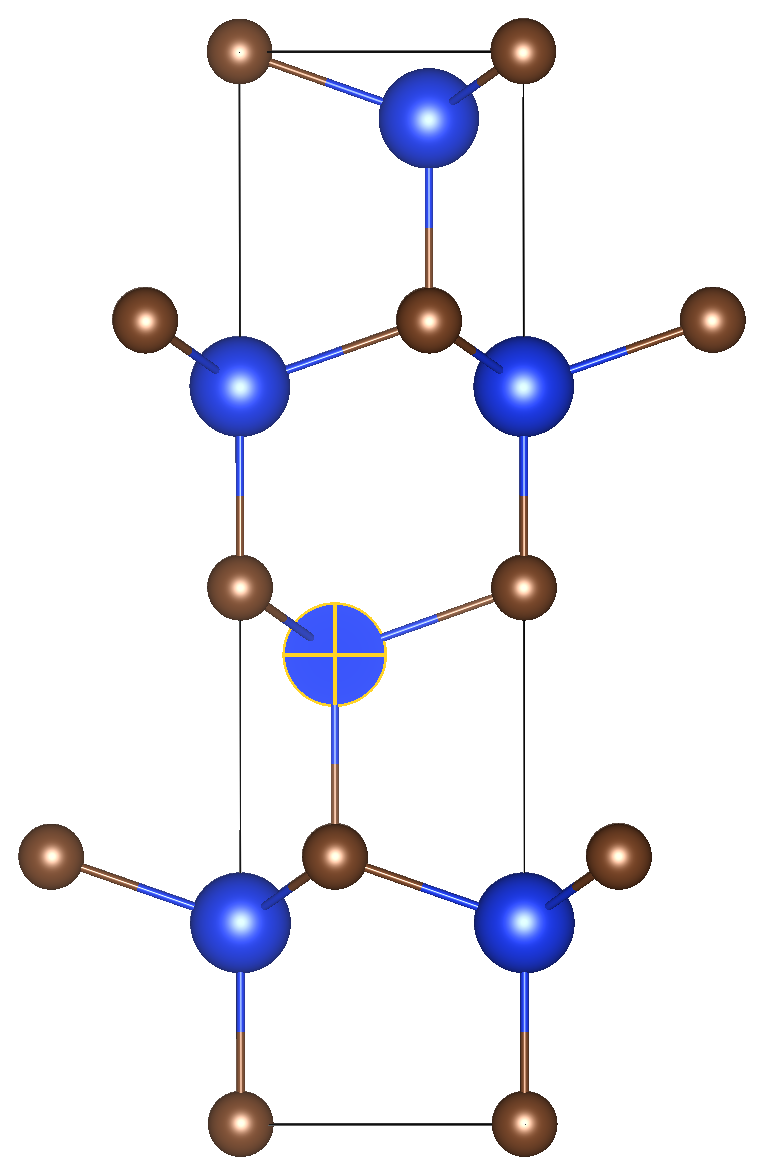}
\end{subfigure}
\hfill
\begin{subfigure}{0.4944\linewidth}
\caption{}
\label{subfig:stirap}
    \includegraphics[width=1.0\linewidth]{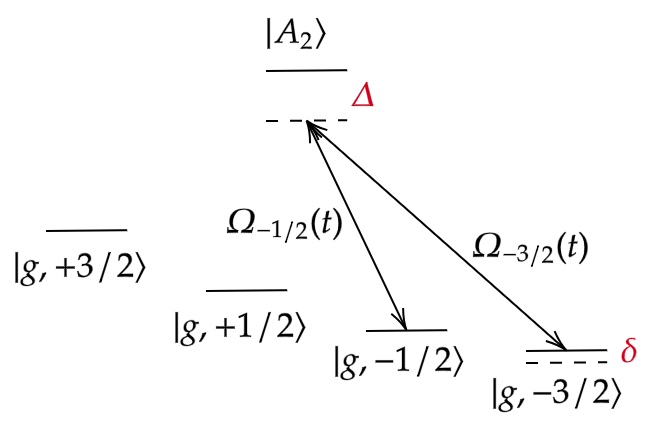}
\end{subfigure}
\caption{(a) 4H-SiC conventional unit cell viewed along the $[010]$ direction in the hexagonal frame of reference, where the highlighted h-site Si atom is removed in order to create a $\mathrm{V}_{\mathrm{Si}}^{-1}$. (b) Graphical depiction of the STIRAP protocol in our case study, allowing for coherent and high fidelity population transfer $\vert\mathrm{g},-\frac{3}{2}\rangle\longrightarrow\vert\mathrm{g},-\frac{1}{2}\rangle$.}
\label{fig:defect and stirap}
\end{figure*}

The concrete case we focus on is SiC, which is widely recognized as a useful material for technological applications, most notably in the microelectronics industry~\cite{elasser2002silicon,eddy2009silicon} and in the Quantum Technology (QT) realm~\cite{castelletto2022silicon,csore2021point}. Among the different structures for the stacking layers known as SiC polytypes, the hexagonal ones (4H-SiC and 6H-SiC) are the most studied in QT due to the availability of high quality samples in the laboratory and the consensus reached in the identification of the ground state and symmetry of the most known defect in them, the negatively charged silicon vacancy ($\mathrm{V}_{\mathrm{Si}}^{-1}$)~\cite{janzen2009silicon}. This defect can be modeled as an electron spin with a spin quartet ground state configuration ($S=\frac{3}{2}$)~\cite{wimbauer1997negatively,ivady2018first,ivady2017identification}, as in Fig.~\ref{fig:defect and stirap}b. Although highly purified 4H-SiC samples are available, in general in non-purified samples there is a $4.7$\% of $^{29}$Si and a $1.1$\% of $^{13}$C nuclear spins whose magnetic, or hyperfine, interaction with the silicon vacancy has been calculated from first principles and measured in experiments~\cite{wimbauer1997negatively,wagner2002ligand,ivady2017identification,son2019ligand}. An appealing property of such defects in the QT realm is their high decoherence time in the millisecond range~\cite{fazio2022decoherence,nagy2018quantum,seo2016quantum}. Furthermore, by being able to be coherently controlled via laser signals and suitably applied magnetic and microwave fields, such defects in SiC are well-known to the QC community~\cite{heremans2016control,wolfowicz2021quantum}. For instance, in Ref.~\cite{nagy2019high} the feasibility of a spin-to-photon interface for quantum information applications involving an hexagonal site (h-site) $\mathrm{V}_{\mathrm{Si}}^{-1}$ in purified 4H-SiC is demonstrated by utilizing a laser pulse resonant to the $A_2$ optical transition, a decay through metastable states and microwave fields in order to selectively initialize the state ($\vert\mathrm{g},-\frac{1}{2}\rangle$), starting from the global ground state ($\vert\mathrm{g},-\frac{3}{2}\rangle$) within the ground state manifold related to an external magnetic field of 92 G. The consequent high fidelity preparation and readout of a quantum state is of paramount importance in the QT field and constitutes one of the reasons why this system has been proposed as a possible hardware for the future quantum computer~\cite{de2021materials} and a material platform for spin-based photonics~\cite{atature2018material}. Our goal in this work is to investigate the possibility of exploiting an additional method for the coherent initialization in $\vert\mathrm{g},-\frac{1}{2}\rangle$ of an h-site $\mathrm{V}_{\mathrm{Si}}^{-1}$ ($\mathrm{V}_1$ center~\cite{nagy2019high,davidsson2022exhaustive}) in non-purified 4H-SiC (see Fig. \ref{fig:defect and stirap}a) based on STIRAP, and study the influence of the $^{29}$Si and $^{13}$C nuclear spin bath in the process. This analysis is appropriate since extensive literature has shown that the presence of a bosonic bath is deleterious for these kinds of control procedures~\cite{ivanov2004effect,ivanov2005spontaneous,vitanov1997population,scala2010stimulated}.


Since in previous papers~\cite{militello2023adiabatic,militello2024adiabatically} interactions between bath spins are not probed, in the present work we examine intra-bath correlations by utilizing the standard one-way STIRAP procedure in order to transfer population in a $\mathrm{V}_{\mathrm{Si}}^{-1}$ in \textit{non-purified} 4H-SiC while probing the influence of the nuclear spin bath in the process. In order to do that we have to apply the STIRAP protocol to a five-level system ($\mathrm{V}_{\mathrm{Si}}^{-1}$ ground state manifold plus one of its optically excited states), which is graphically depicted in Fig. \ref{subfig:stirap}. As can be seen in Fig. 1b of Ref.~\cite{nagy2019high}, in the single-particle picture, a transition to the excited state manifold is obtained through an electronic transition within the bandgap in which the electron cannot flip due to existing selection rules, hence our choice for $\vert A_2\rangle\equiv\vert\mathrm{e},-\frac{3}{2}\rangle$ as intermediate state. Therefore, we propose a procedure allowing for the coherent initialization of the first excited state of the ground state manifold starting from the global ground state ($\vert\mathrm{g},-\frac{3}{2}\rangle$), but including the bath in our analysis. From an experimental perspective, we assume that optical pumping is used to populate only the global ground state prior to the proposed STIRAP protocol and the populations could be read out by applying an optical readout laser pulse resonant with the $A_2$ transition, as in Ref.~\cite{nagy2019high}. This accomplishes the same result by additionally extracting useful indirect information on the nuclear spin bath in the meantime. An arising issue concerns the decoherence time of the $\mathrm{V}_{\mathrm{Si}}^{-1}$ being of $\sim10\hspace{0.1cm}\mathrm{\mu s}$ after free evolution~\cite{fazio2022decoherence}, while STIRAP being a slow process in order to ensure adiabaticity. This is resolved by exploiting the framework proposed by Dogra et al.~\cite{dogra2022perfect} for a perfect STIRAP with imperfect finite-time Gaussian pulses. We demonstrate that in this way we are able to perfectly transfer population in our five-level system in a time period $T\sim1\hspace{0.1cm}\mathrm{\mu s}$ (Sec. \ref{subsec:findings}).

Finally, another issue regards the multiplicity of the bath containing $\sim1500$ nuclear spins inside a spherical bath with a radius of $\sim5\hspace{0.1cm}\mathrm{nm}$~\cite{seo2016quantum}. In order to solve this issue, also in view of considering intra-bath interactions and consequent correlations, we resort to the use of Cluster Correlation Expansion (CCE) theory, introduced in Yang et al.~\cite{yang2008quantum} and successfully applied to a condensed matter system similar to ours soon thereafter~\cite{balian2014quantum,balian2015keeping,seo2016quantum,fazio2022decoherence}. Although in its initial conception CCE theory was designed to calculate the off-diagonal components of the density matrix of spin systems randomly interacting with fermionic spin baths, a later generalization by Yang et al.~\cite{yang2020longitudinal} allowed to evaluate also the diagonal components, or populations. Therefore, the generalized CCE (gCCE) theory was born, which includes the central electron spin in each cluster and has been already recently applied to study clock transitions of a divacancy in 4H-SiC~\cite{onizhuk2021probing}.

The paper is organized as follows: in Sec. \ref{sec:model_methods} we introduce the Hamiltonian model for a few-state system driven by laser pulses (\ref{subsec:model}) and magnetically interacting with a bath of nuclear spins, as well as the methods to study this situation based on gCCE (\ref{subsec:methods}). In Sec. \ref{sec:application} we specialize the general model to the level structure of a SiC (\ref{subsec:ham_manip}) and then present the results of the relevant numerical simulations (\ref{subsec:findings}). 
Finally, in Sec. \ref{sec:discuss} we provide additional insights and a summary of our results. We conclude with an appendix regarding the convergence properties of the procedure (Appendix \ref{app:A}).

\section{\label{sec:model_methods}Model and methods}

\subsection{\label{subsec:model}Model}

From a modeling perspective the system we are concerned with is a five-state system interacting with a fermionic bath of two-level systems: three states are properly involved in the STIRAP scheme, while two more states are involved in the interaction with a spin bath. States $\ket{a_1}$ and $\ket{a_2}$ are the initial and target state, while $\ket{b}$ is the auxiliary state coupled to the first two through suitable pulse usually referred to as pump and Stokes. Finally, the four states $\ket{a_1}$, $\ket{a_2}$, $\ket{c_1}$ and $\ket{c_2}$ are in general connected through the interaction with the bath (in Fig.~\ref{fig:generalscheme} it is represented the scheme involving both the pulses and the interaction with the bath). The relevant Hamiltonian is expressible as follows:
\begin{eqnarray}
\nonumber
\mathcal{H}&=& \sum_{k=1,2} \omega_{a_k} \ket{a_k}\bra{a_k}  
+ \omega_b \ket{b}\bra{b}  
+ \sum_{k=1,2} \omega_{c_k} \ket{c_k}\bra{c_k}  
\\
\nonumber
&+& \sum_{k=1,2} \left( \Omega_k (t) e^{-\ii \tilde{\omega}_k t} + \mathrm{h.c.} \right) \left( \ket{a_k}\bra{b} + \mathrm{h.c.} \right)
\\
&-&\sum_{i=1}^N\omega_{I_i}I_{iz} +\mathcal{H}_{BI} 
+ \sum_\alpha X_\alpha Y_\alpha,
\label{eq:three_level_ham}
\end{eqnarray}
where $X_\alpha$ (acting only inside the quadruplet $\{\ket{a_1}, \ket{a_2}, \ket{c_1}, \ket{c_2}\}$) and $Y_\alpha$ are suitable system and bath operators describing the system-environment interaction, while $\Omega_k$ and $\tilde{\omega}_k\equiv \omega_{a_k} - \omega_{c_k}$ describe the action of STIRAP pulses. The pulses' profile described by the $\Omega_k$ is usually Gaussian. The term  $\mathcal{H}_{BI}$ describes the interaction between the spins of the bath. Finally, $\omega_{I_i}=\gamma_i B$ is the Larmor frequency associated to the $i$-th spin of the bath, with $\gamma_i$ the gyromagnetic ratio of the i-th nuclear spin.

\subsection{\label{subsec:methods}Methods}

In this subsection we share the methods we have utilized in order to study our system and solve the dynamics guided by Eq. \ref{eq:three_level_ham}.

\begin{figure*}
\begin{center}
    \includegraphics[width=1.0\textwidth]{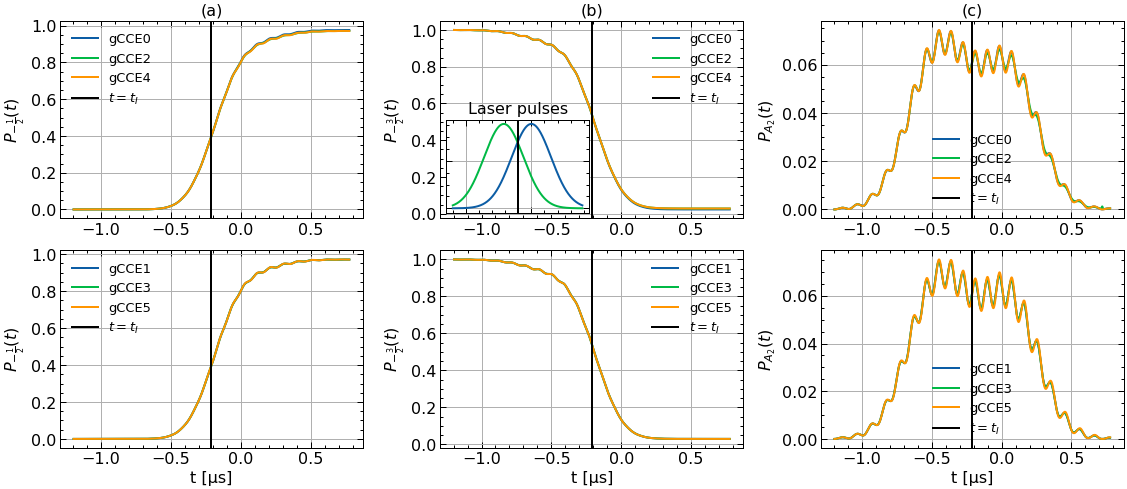}
    \caption{Populations of the electron spin's $\vert\mathrm{g},-1/2\rangle$ (left column), $\vert\mathrm{g},-3/2\rangle$ (middle column) and $\vert A_2\rangle$ (right column) states as a function of time, for different orders of gCCE. The bath is spherical with a radius $R=2$ nm ($\sim100$ nuclear spins), the external magnetic field is $B=10$ mT and the detuning is $\Delta=60$ MHz. Wherever it is not otherwise stated the remaining parameters are chosen from the set corresponding to a perfect STIRAP (Subsec. \ref{subsec:findings}). The time dependency of the pump (blue) and Stokes (green) laser pulses (in MHz) given in Eqs. \ref{eq:pump} and \ref{eq:Stokes} is shown in the inset ($\Omega=35.6$ MHz).}
\label{fig:gCCE convergence}
\end{center}
\end{figure*}

In order to simulate the dynamics guided by the Hamiltonian in Eq. \ref{eq:three_level_ham} we chose first an initial state for the nuclear spin bath. The nuclear spins were modeled as spin-1/2 with a temperature such that $k_{\mathrm{B}}T_{\mathcal{B}}\gg\omega_{I_i}$ for all $i=1,\ldots,N$ (with $k_{\mathrm{B}}$ the Boltzmann's constant) and their overall initial state, which is a thermal state at high temperature, is well approximated by a purely mixed one,
\begin{equation}
    \vert\mathcal{B}(0)\rangle=\bigotimes_{i=1}^N\frac{I_i}{2}, \label{eq:bath_init_state}    
\end{equation}
where $I_i$ is the i-th nuclear spin's identity operator. In the ideal case the system would stay in the dark state~\cite{kumar2016stimulated,dogra2022perfect},
\begin{equation}
    \vert\mathrm{D}\rangle=\cos\theta(t)\big\vert a_1\big\rangle-\sin\theta(t)\big\vert a_2\big\rangle, \label{eq:def_init_state}    
\end{equation}
for the whole duration of the time evolution, with an infidelity $\epsilon=\sin\theta_i=\cos\theta_f=0.01$ and a fidelity $\sqrt{1-\epsilon^2}$ expressing the efficiency of our protocol. Furthermore, $\theta(t)$ is the mixing angle, being defined as $\theta(t)=\tan^{-1}(\Omega_1(t)/\Omega_2(t))$, where $\Omega_1(t)$ and $\Omega_2(t)$ are the Gaussian amplitudes of the pump and Stokes laser pulses, respectively, given by~\cite{dogra2022perfect} (see inset of Fig. \ref{fig:gCCE convergence})
\begin{eqnarray}
    \Omega_1(t)=\Omega\hspace{0.05cm}e^{-t^2/2\sigma^2}, \label{eq:pump} \\
    \Omega_2(t)=\Omega\hspace{0.05cm}e^{-(t-t_s)^2/2\sigma^2}. \label{eq:Stokes}
\end{eqnarray}
Here, $\Omega$ is the common laser pulses' amplitude, $\sigma$ is the width of the pulses and $t_s$ is the time shift between the pulses. We also used a dimensionless shift as $s=t_s/\sigma$. The dynamics was analyzed during a finite time from an initial time $t_i=-(n_t-s)\sigma$ to a final time $t_f=n_t\sigma$, where $n_t$ is a real number,
\begin{equation}
    n_t=\frac{1}{s}\ln\frac{\epsilon}{\sqrt{1-\epsilon^2}}+\frac{s}{2}. \label{eq:n_t}    
\end{equation}
Once a value for $\sigma$ was chosen, the amplitude $\Omega$ was fixed by demanding that it satisfies the contingent adiabaticity criterion from Ref.~\cite{dogra2022perfect},
\begin{equation}
    \sigma\Omega\gtrsim\frac{-s e^{s^2/8}}{2(2-\sqrt{2})}, \label{eq:cont_adiab_crit}    
\end{equation}
and the global adiabatic condition from Ref.~\cite{kumar2016stimulated},
\begin{equation}
    \sigma\Omega\gg\frac{\sqrt{\pi}}{4},    
\end{equation}
which has meant $\Omega=35.6$ MHz.

In order to analyze the effects of the environment, even in the presence of possible intra-bath interactions, we have exploited the gCCE theory. Whenever a bath was present (going beyond gCCE0, i.e. no system-bath interaction), the generic population was calculated by means of a product expansion in cluster-correlation terms of increasing order, i.e. increasing number of fully-interacting nuclear spins inside the clusters~\cite{yang2008quantum,yang2020longitudinal},
\begin{equation}
    P(t)=\Tilde{P}_{\lbrace0\rbrace}\prod_i\Tilde{P}_{\lbrace i\rbrace}\prod_{i,j}\Tilde{P}_{\lbrace i,j\rbrace}\cdots. \label{eq:gCCE_prod_exp}    
\end{equation}
The product expansion is justified by assuming that the clusters are non-interacting with one another. The contribution to the total population from the generic cluster $C$ is recursively defined as
\begin{equation}
    \Tilde{P}_{\lbrace C\rbrace}=\frac{P_{\lbrace C\rbrace}(t)}{\prod\limits_{C'\subset C}\Tilde{P}_{\lbrace C'\rbrace}}, \label{eq:gCCE_contr_clust_C}    
\end{equation}
where the population including the nuclear spins in cluster $C$ is calculated according to its definition, i.e.
\begin{equation}
    P_{\lbrace C\rbrace}^{a_k}(t)=\frac{\mathrm{tr}\left\lbrace\rho_C(t)\big\vert a_k\big\rangle\big\langle a_k\big\vert\right\rbrace}{\mathrm{tr}\left\lbrace\rho_C(0)\big\vert a_k\big\rangle\big\langle a_k\big\vert\right\rbrace}. \label{eq:pop_clust_C}    
\end{equation}
The definition in Eq. \ref{eq:gCCE_contr_clust_C} is due to the fact that for instance the third order truncation of \ref{eq:gCCE_prod_exp} gives the exact formula for the population whenever there are only three nuclear spins in the bath. In Eq. \ref{eq:pop_clust_C}, $\rho_C(t)=U_C(t)\rho_C(0)U_C^\dagger(t)$ and
\begin{equation}
    U_C(t)=\mathcal{T}e^{-\ii\int_{t_i}^t\mathcal{H}_C(\tau)\mathrm{d}\tau}, \label{eq:time_evol_oper}    
\end{equation}

\noindent in which $\mathcal{H}_C(\tau)$ is the Hamiltonian given in Eq. \ref{eq:three_level_ham} where the pump and Stokes amplitudes are time-dependent (Eqs. \ref{eq:pump} and \ref{eq:Stokes}) and only the nuclear spins inside cluster $C$ are considered. As for the bath itself, the nuclear spins were randomly generated in the right abundance known from experiments and put in the lattice until a given radius $R$ was reached (the bath has a spherical shape). Furthermore, a distance $r$, often called nuclear spin connectivity, was introduced in order to set a maximum distance beyond which a pair of nuclear spins is no longer interacting. These two parameters offer a constraint on the coupling constants between the system and the bath and among bath spins in Eq. \ref{eq:three_level_ham}, respectively. Finally, one last parameter $n$, the number of realizations of the bath, was added in order to perform a statistical sampling of the random bath-generating procedure. For each of the $n$ realizations the associated populations were calculated via Eq. \ref{eq:gCCE_prod_exp} and in the end a mean was carried out in order to obtain the final populations. In Appendix \ref{app:A} the convergence properties of the bath numerical parameters can be found for a particular concrete case, a $\mathrm{V}_{\mathrm{Si}}^{-1}$ in 4H-SiC. \\
\hspace*{0.5cm}Finally, whenever the bath was absent (gCCE0) the populations were calculated directly through Eq. \ref{eq:pop_clust_C}.

\section{\label{sec:application}Application to a $\mathrm{V}_{\mathrm{Si}}^{-1}$ in 4H-SiC}

\subsection{\label{subsec:ham_manip}Hamiltonian manipulation}

In this subsection we specialize our analysis to a particular case, a $\mathrm{V}_{\mathrm{Si}}^{-1}$ in 4H-SiC, which is useful in that its physical parameters, like the ones appearing in Eq. \ref{eq:three_level_ham}, are known in the literature~\cite{soykal2016silicon}. Its four-level structure, depicted in Fig. \ref{subfig:stirap}, is due to the lowest energy electronic configuration involving five valence electrons, two in a single-particle state having spin up and down and three in another one having spin up~\cite{ivady2018first}. The system plus bath Hamiltonian is again most conveniently expressed in the electron spin's $S_z$ operator's eigenbasis $\lbrace\vert3/2\rangle,\hspace{0.1cm}\vert1/2\rangle,\hspace{0.1cm}\vert-1/2\rangle,\hspace{0.1cm}\vert-3/2\rangle\rbrace$. This time, as auxiliary state to apply STIRAP we chose $\vert\mathrm{e},-3/2\rangle\equiv\vert A_2\rangle$, which is excited by means of an optical transition of 1.44 eV with $A_2$ symmetry, known as V1 line in the literature~\cite{nagy2019high,davidsson2022exhaustive}. Therefore~\cite{sakuldee2019characterization},
\small{
\begin{equation}
\begin{split}
    \mathcal{H}&=\big(-2D-\omega_e\big)\Big\vert\mathrm{g},-\frac{1}{2}\Big\rangle\Big\langle\mathrm{g},-\frac{1}{2}\Big\vert \\
    &+\big(-2D-2\omega_e\big)\Big\vert\mathrm{g},\frac{1}{2}\Big\rangle\Big\langle\mathrm{g},\frac{1}{2}\Big\vert \\
    &-3\omega_e\Big\vert\mathrm{g},\frac{3}{2}\Big\rangle\Big\langle\mathrm{g},\frac{3}{2}\Big\vert+\omega_{A_2}\big\vert A_2\big\rangle\big\langle A_2\big\vert-\sum_{i=1}^N\omega_{I_i}I_{iz} \\
    &+\sum_{i=1}^N\sum_{\alpha,\beta}S_\alpha A^i_{\alpha\beta}I_{i\beta}+\sum_{i<j=1}^N\sum_{\alpha,\beta}I_{i\alpha}B^{ij}_{\alpha\beta}I_{j\beta}, \label{eq:init_ham}
\end{split}
\end{equation}}\normalsize{where} 
$D$ is the Zero-Field Splitting (ZFS) tensor's longitudinal component ($D=2.5$ MHz for a $\mathrm{V}_{\mathrm{Si}}^{-1}$ in 4H-SiC~\cite{son2019ligand,nagy2019high}), $\omega_e=\gamma_eB$ is the Larmor frequency of the electron spin ($\gamma_e$ being its gyromagnetic ratio), while $A^i_{\alpha\beta}$ and $B^{ij}_{\alpha\beta}$ are the hyperfine and dipolar tensors' components. These components are in this case accurately described in the dipolar approximation~\cite{golding1982evaluation},
\begin{equation}
    A^i_{\alpha\beta}=\frac{\mu_0\gamma_i\gamma_e}{4\pi r_i^3}\left(\delta_{\alpha\beta}-\frac{3r_{i\alpha}r_{i\beta}}{r_i^2}\right) \label{eq:dip_hyp_tens}
\end{equation}
and
\begin{equation}
    B^{ij}_{\alpha\beta}=\frac{\mu_0\gamma_i\gamma_j}{4\pi r_{ij}^3}\left(\delta_{\alpha\beta}-\frac{3r_{ij\alpha}r_{ij\beta}}{r_{ij}^2}\right). \label{eq:dip_tens}
\end{equation}
Here $\Vec{r}_i$ is the position vector locating the i-th nuclear spin with respect to the electron spin at the origin of a cartesian reference frame and $\mu_0$ is the vacuum magnetic permeability. Furthermore, $\Vec{r}_{ij}=\Vec{r}_i-\Vec{r}_j$ is the position vector pointing from the j-th to the i-th nuclear spin, $r_{ij}$ is its modulus and $r_{ij\alpha}$ its $\alpha$-component. The energy of the $\vert\mathrm{g},-3/2\rangle$ state has been chosen to be the zero of energy.

In order to correctly apply STIRAP to trigger a population exchange between $\vert\mathrm{g},-3/2\rangle$ and $\vert\mathrm{g},-1/2\rangle$, we must apply two high frequency laser pulses almost resonant with the $A_2$ transition having a Gaussian-shaped envelope, as in Ref.~\cite{dogra2022perfect}. The pulses' amplitudes are chosen equal and the specific value, along with other important pulse parameters like width and distance between them, is singled out by utilizing the procedure layed out in Ref.~\cite{dogra2022perfect} in order to obtain perfect STIRAP with imperfect finite-time pulses. This is helpful in order for the population transfer to occur in a time shorter than the decoherence time of the system ($\sim10\hspace{0.1cm}\mu\mathrm{s}$)~\cite{fazio2022decoherence,nagy2018quantum,seo2016quantum}. In general, the action of the pump and Stokes pulses can be described by the following generalized Rabi-type perturbation Hamiltonian,
\begin{equation}
\begin{split}
    \mathcal{H}_1=&-\left(\Omega_\mathrm{P}e^{-\ii\omega_\mathrm{P}t} + \mathrm{h.c.}\right) \left( \Big\vert A_2\Big\rangle\Big\langle\mathrm{g},-\frac{3}{2}\Big\vert+\mathrm{h.c.} \right) \\
    &-\left(\Omega_\mathrm{S}e^{-\ii\omega_\mathrm{S}t}+  \mathrm{h.c.} \right)
    \left( \Big\vert A_2\Big\rangle\Big\langle\mathrm{g},-\frac{1}{2}\Big\vert+\mathrm{h.c.} \right), \label{eq:pump_stokes_pulses}
\end{split}    
\end{equation}
where $\Omega_{\mathrm{P/S}}$ are the laser's Rabi frequencies (amplitudes) associated to the pump and Stokes signals, respectively, whereas $\omega_{\mathrm{P/S}}$ are the frequencies of the pump and Stokes signals. The Hamiltonian of Eq. \ref{eq:init_ham} plus the pulses in Eq. \ref{eq:pump_stokes_pulses} share the same form as that in Eq. \ref{eq:three_level_ham} provided $\vert\mathrm{g},-3/2\rangle=\vert a_1\rangle$, $\vert\mathrm{g},-1/2\rangle=\vert a_2\rangle$, $\vert A_2\rangle=\vert b\rangle$, $\vert\mathrm{g},3/2\rangle=\vert c_1\rangle$, $\vert\mathrm{g},1/2\rangle=\vert c_2\rangle$, $\sum_{i=1}^N\sum_\beta A^i_{\alpha\beta}I_{i\beta}=Y_\alpha$, $\sum_{i<j=1}^N\sum_{\alpha,\beta}I_{i\alpha}B^{ij}_{\alpha\beta}I_{j\beta}=\mathcal{H}_{BI}$, $\Omega_{\mathrm{P/S}}=\Omega_{1/2}$ and $\omega_{\mathrm{P/S}}=\Tilde{\omega}_{1/2}$. Moreover, $X_\alpha = S_\alpha$ (the components of the spin operator in the quadruplet), which implies that only three of the six couplings in the scheme of Fig. \ref{fig:generalscheme} are active: $\ket{a_1}\leftrightarrow\ket{a_2}$, $\ket{a_2}\leftrightarrow\ket{c_2}$, $\ket{c_2}\leftrightarrow\ket{c_1}$. The counter-rotating terms are usually negligible whenever an almost resonant pump laser pulse is chosen, due to the fact that $\omega_{A_2}-\omega_{\mathrm{P}}\ll\omega_{A_2}+\omega_{\mathrm{P}}$ and the fast rotating part with frequency $\omega_{A_2}+\omega_{\mathrm{P}}$ averages out to zero. In order to correctly perform the Rotating Wave Approximation (RWA), we switch to the rotating frame through the unitary transformation $U=e^{\ii\omega_{A_2}t\vert A_2\rangle\langle A_2\vert}$ , discard the fast rotating terms and go back to the original picture. After that and by taking the amplitudes as real functions, the perturbation Hamiltonian becomes
\begin{equation}
\begin{split}
    \mathcal{H}_1=&-\Omega_{\mathrm{P}}\left(e^{-\ii\omega_{\mathrm{P}}t}\Big\vert A_2\Big\rangle\Big\langle\mathrm{g},-\frac{3}{2}\Big\vert+\mathrm{h.c.}\right) \\
    &-\Omega_{\mathrm{S}}\left(e^{-\ii\omega_{\mathrm{S}}t}\Big\vert A_2\Big\rangle\Big\langle\mathrm{g},-\frac{1}{2}\Big\vert+\mathrm{h.c.}\right). \label{eq:pump_stokes_pulses_rwa}
\end{split}    
\end{equation}
Finally, after having applied the transformations $e^{-\ii(\omega_{\mathrm{P}}t + \pi)\vert\mathrm{g},-\frac{3}{2}\rangle\langle\mathrm{g},-\frac{3}{2}\vert} e^{-\ii(\omega_{\mathrm{S}}t+\pi)\vert\mathrm{g},-\frac{1}{2}\rangle\langle\mathrm{g},-\frac{1}{2}\vert}$, our final Hamiltonian is
\footnotesize{
\begin{equation}
\begin{split}
    \tilde{\mathcal{H}}&=\big(-2D-\omega_e-\Delta_{\mathrm{pul}}\big)\Big\vert\mathrm{g},-\frac{1}{2}\Big\rangle\Big\langle\mathrm{g},-\frac{1}{2}\Big\vert \\
    &+\big(-2D-2\omega_e+\Delta-\omega_{A_2}\big)\Big\vert\mathrm{g},\frac{1}{2}\Big\rangle\Big\langle\mathrm{g},\frac{1}{2}\Big\vert \\
    &+\big(-3\omega_e+\Delta-\omega_{A_2}\big)\Big\vert\mathrm{g},\frac{3}{2}\Big\rangle\Big\langle\mathrm{g},\frac{3}{2}\Big\vert \\
    &+\Delta\big\vert A_2\big\rangle\big\langle A_2\big\vert-\sum_{i=1}^N\omega_{I_i}I_{iz}+\sum_{i=1}^N\sum_\alpha S_z A^i_{z\alpha}I_{i\alpha} \\
    &+\sum_{i=1}^N\sum_\alpha\left[\frac{\sqrt{3}}{2}\Big\vert\mathrm{g},-\frac{3}{2}\Big\rangle\Big\langle\mathrm{g},-\frac{1}{2}\Big\vert e^{-\ii\Delta_{\mathrm{pul}}t}(A^i_{x\alpha}+iA^i_{y\alpha})+\mathrm{h.c.}\right]I_{i\alpha} \\
    &+\sum_{i<j=1}^N\sum_{\alpha,\beta}I_{i\alpha}B^{ij}_{\alpha\beta}I_{j\beta} \\
    &+\frac{\Omega_{\mathrm{P}}}{2}\left(\Big\vert A_2\Big\rangle\Big\langle\mathrm{g},-\frac{3}{2}\Big\vert+\mathrm{h.c.}\right)+\frac{\Omega_{\mathrm{S}}}{2}\left(\Big\vert A_2\Big\rangle\Big\langle\mathrm{g},-\frac{1}{2}\Big\vert+\mathrm{h.c.}\right). \label{eq:final_ham}
\end{split}
\end{equation}}

\normalsize{Here} the $\vert\mathrm{g},-3/2\rangle$ state eigenenergy is again considered as the zero of energy, $\Delta_{\mathrm{pul}}=\omega_{\mathrm{P}}-\omega_{\mathrm{S}}$ is the difference between the pump and Stokes frequencies, and $\Delta=\omega_{A_2}-\omega_{\mathrm{P}}$ is the detuning with the optical $A_2$ transition. In order for the STIRAP process to be successful we have to allow for the degeneracy of the $\vert\mathrm{g},-3/2\rangle$ and $\vert\mathrm{g},-1/2\rangle$ states, which is true if $\Delta_{\mathrm{pul}}$ is chosen as $\Delta_{\mathrm{pul}}=-2D-\omega_e$. This is equivalent to the two-photon resonance condition ($\delta=0$) met in ordinary STIRAP being applied to genuine three-level systems~\cite{bergmann2019roadmap}. Finally, notice that even though the electron spin is a five-level system, if prepared in its ground state and subjected to the dynamics guided by Eq. \ref{eq:final_ham}, where no transitions can occur towards the $\vert\mathrm{g},3/2\rangle$ and $\vert\mathrm{g},1/2\rangle$ states, it effectively becomes a three-level system. Additionally, these transitions are not allowed indirectly through the bath's intermediary action due to the enormous mismatch between the associated transition frequencies and the nuclear spins' transition frequency ($10^8$ MHz compared to $10^{-1}$ MHz for $B\sim10$ mT, respectively). Therefore, in the remainder of the paper we plot only the populations of the $\vert\mathrm{g},-3/2\rangle$, $\vert\mathrm{g},-1/2\rangle$ and $\vert A_2\rangle$ states, the others being always unpopulated. However, the presence of the spin bath (particularly in the nearest-neighbor sites) could compromise the STIRAP process by means of flip-flop transitions causing the electron spin to lower its energy by remaining in the excited state manifold and avoiding the aforementioned energy mismatch. This could happen, e.g., with the $\vert\mathrm{e},-\frac{3}{2}\rangle\longrightarrow\vert\mathrm{e},-\frac{1}{2}\rangle$ transition, which is magnetic dipole allowed~\cite{nagy2019high}.

\subsection{\label{subsec:findings}Findings}

\begin{figure}
\centering
\begin{subfigure}{0.4944\linewidth}
\caption{}
\label{subfig:gCCE0_Delta50}
    \includegraphics[width=1.0\linewidth]{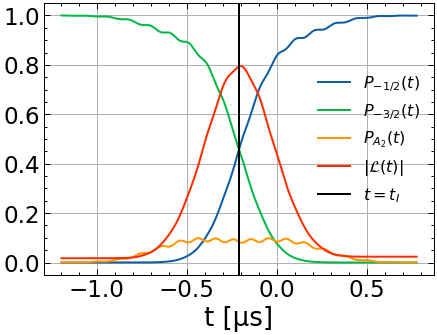}
\end{subfigure}
\hfill
\begin{subfigure}{0.4944\linewidth}
\caption{}
\label{subfig:gCCE0_Delta60}
    \includegraphics[width=1.0\linewidth]{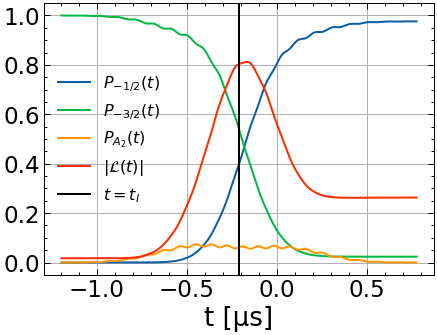}
\end{subfigure} \\
\begin{subfigure}{0.4944\linewidth}
\caption{}
\label{subfig:gCCE1_Delta50}
    \includegraphics[width=1.0\linewidth]{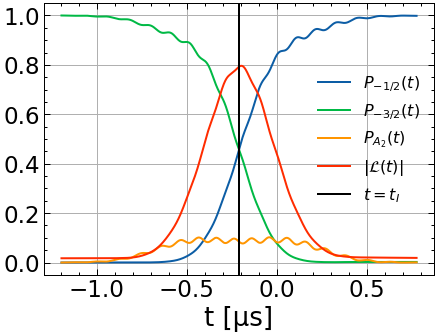}
\end{subfigure}
\begin{subfigure}{0.4944\linewidth}
\caption{}
\label{subfig:gCCE1_Delta60}
    \includegraphics[width=1.0\linewidth]{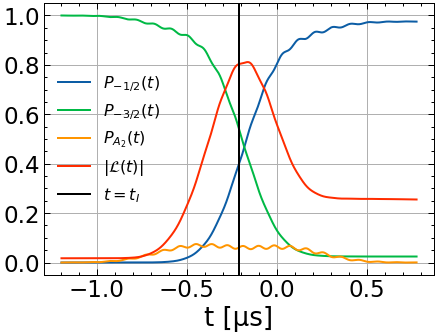}
\end{subfigure}
\caption{Graphs showing the populations in each state of the ground state manifold involved in the STIRAP process and the coherence modulus of a $\mathrm{V}_{\mathrm{Si}}^{-1}$ in 4H-SiC, for $B=10$ mT. In (a) and (b) the bath is absent (gCCE0), whereas in (c) and (d) it is present (gCCE1). For (a) and (c) $\Delta=50$ MHz, whereas for (b) and (d) $\Delta=60$ MHz. In all cases $\Omega=35.6$ MHz.}
\label{fig:gCCE0 and gCCE1}
\end{figure}

In order to understand the system's behavior and how it is affected by the bath, in this subsection we turn to simulations by 
evaluating the dynamics by means of an approximation through gCCE theory (see Subsec. \ref{subsec:methods} for additional information).

A useful initial insight comes from the order of the gCCE at which convergence is achieved. Here convergence is reached whenever a change in the parameter does not lead to a qualitative change in the populations. The gCCE order is the number of fully-interacting nuclear spins inside the biggest clusters of the expansion, thus being linked to the effect many-body correlations within the bath have on the electron spin~\cite{yang2008quantum,yang2009quantum}. Therefore, the converged value of this parameter allows us to infer how the interactions between nuclear spins affect the electron spin and the number of fully-interacting ones which maximize this effect. See, e.g., Figs. \ref{fig:gCCE convergence}a, \ref{fig:gCCE convergence}b and \ref{fig:gCCE convergence}c, in which we separately plot the populations of the electron spin's $\vert\mathrm{g},-1/2\rangle$, $\vert\mathrm{g},-3/2\rangle$ and $\vert A_2\rangle$ states as a function of time, for different orders of gCCE (the zeroth order depicting a situation where the bath is absent). Notice that $P_{-1/2}$ correctly increases while $P_{-3/2}$ decreases, and the population exchange is centered at the intermediate time $t=t_I$ where the Stokes laser pulse is equal to the pump one~\cite{dogra2022perfect} (see the inset in Fig. \ref{fig:gCCE convergence}b). Such time instant is chosen as the zero of the time axis for symmetry reasons, and the usual positive values for $t$ may be obtained via a simple rigid shift of the axis. In this choice we follow the example of Dogra et al.~\cite{dogra2022perfect}. All populations, and in particular $P_{A_2}$, show a wiggly behavior in addition to the overall trend, which is due to $\Delta_{\mathrm{pul}}$ in Eq. \ref{eq:final_ham} or to the laser frequencies $\omega_{\mathrm{P}}$ and $\omega_{\mathrm{S}}$. As expected, the $\vert A_2\rangle$ state is only ever populated by a small fraction at the intermediate point of the dynamics. However, an outcome of our analysis is the fact that the presence of a bath does not modify the electron spin's populations, which is witnessed by the dynamics being described to a high degree of accuracy already at gCCE0. Therefore, many-body interactions between the nuclear spins interacting with the electron spin at the origin do not add new population effects in the full description of the STIRAP process. One possible explanation for this involves the hyperfine and dipolar tensors' components $A^i_{\alpha\beta}$ and $B^{ij}_{\alpha\beta}$. Since $\Omega\gg A^i_{\alpha\beta},\hspace{0.05cm}B^{ij}_{\alpha\beta}$ ($\Omega\sim1-100\hspace{0.1cm}\mathrm{MHz}$, $A^i_{\alpha\beta}\sim\mathrm{Hz}$ and $B^{ij}_{\alpha\beta}\sim10^{-5}\hspace{0.1cm}\mathrm{Hz}$), the laser part of the Hamiltonian dominates on the terms describing the bath. Therefore, in the investigated duration of the process of $1.97\hspace{0.1cm}\mu\mathrm{s}$ the effect of one- or more-body interactions on the electron spin is not appreciable. In the remainder of our work all simulations will be carried out at the gCCE1 level.

At this point, we are able to show that a perfect STIRAP in a finite time is indeed possible for the studied system, as can be seen in Fig. \ref{fig:gCCE0 and gCCE1}. In particular, Figs. \ref{fig:gCCE0 and gCCE1}a and \ref{fig:gCCE0 and gCCE1}c allow us to claim that the set of parameters giving rise to a perfect population transfer is the following, $\lbrace\epsilon=0.01$, $\sigma=0.3\hspace{0.1cm}\mu\mathrm{s}$, $s=-1.4$, $\Delta_{\mathrm{pul}}=-2D-\omega_e$, $\Delta=50$ MHz, $\Omega=35.6$ MHz$\rbrace$. Here $\epsilon$ is the STIRAP infidelity~\cite{dogra2022perfect} defined under Eq. \ref{eq:def_init_state}, $\sigma$ is the width of the pulses, whereas $s$ and $\Omega$ are their separation and amplitude ($\Omega_{\mathrm{P}}=\Omega_{\mathrm{S}}\equiv\Omega$), respectively. A negative $s$ means a counter-intuitive set of laser pulses, which is often the more convenient choice in the literature~\cite{malinovsky1997simple,sola1999optimal,malinovsky2004quantum}. By changing these parameters one wanders astray from the perfect scenario, as displayed in Figs. \ref{fig:gCCE0 and gCCE1}b and \ref{fig:gCCE0 and gCCE1}d, where a change in $\Delta$ causes the populations to not equal at $t=t_I$, thus preventing the complete $100$\% transfer of population from $\vert\mathrm{g},-3/2\rangle$ to $\vert\mathrm{g},-1/2\rangle$. This is due to the fact that the new value for the detuning is not counterbalanced by an increased value of the pulse amplitude $\Omega$, which is no longer sufficient to ensure the correct realization of the process. As a further proof of that, we plot also the absolute value of the electron spin's coherence, or the off-diagonal component of its density matrix. As expected, it reaches a maximum where the mixture of $\vert\mathrm{g},-3/2\rangle$ and $\vert\mathrm{g},-1/2\rangle$ is highest, at $t=t_I$, and decays afterwards to a fixed nonzero value. This is equal to the initial coherence for $\Delta=50$ MHz, while being much larger for $\Delta=60$ MHz, which is a sign of increased mixture and consequently decreased efficiency of STIRAP. However, a $16.7$\% reduction in detuning is only responsible of a $2.5$\% reduction in STIRAP efficiency, quantified by the final population $P_{-1/2}(t_f)$. Therefore, our proposed procedure is also stable with respect to variations from the aforementioned perfect set of parameter values.

\begin{figure}
\begin{center}
    \includegraphics[width=0.4944\textwidth]{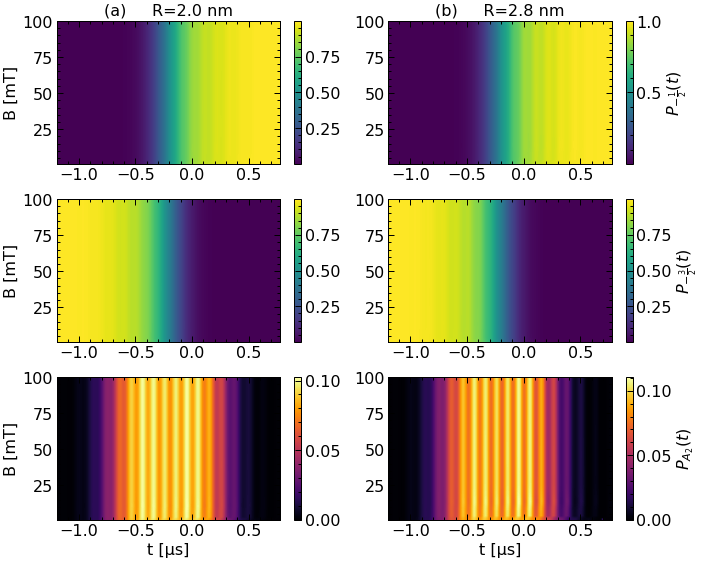}
    \caption{Filled contour plot showing the gCCE1 populations of the electron spin's $\vert\mathrm{g},-1/2\rangle$, $\vert\mathrm{g},-3/2\rangle$ and $\vert A_2\rangle$ states as a function of time and magnetic field. The bath is spherical with a radius $R=2.0$ nm (a) or $R=2.8$ nm (b). In all cases $\Delta=50$ MHz and $\Omega=35.6$ MHz.}
    \label{fig:gCCE1 var B}
\end{center}
\end{figure}

\begin{figure}
\begin{center}
    \includegraphics[width=0.4944\textwidth]{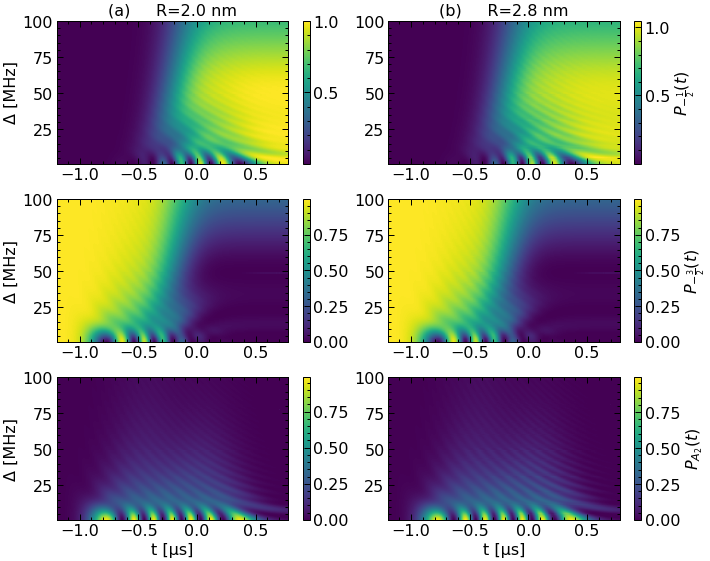}
    \caption{Filled contour plot showing the gCCE1 populations of the electron spin's $\vert\mathrm{g},-1/2\rangle$, $\vert\mathrm{g},-3/2\rangle$ and $\vert A_2\rangle$ states as a function of time and detuning $\Delta$. The bath is spherical with a radius $R=2.0$ nm (a) or $R=2.8$ nm (b). In all cases $B=10$ mT and $\Omega=35.6$ MHz.}
    \label{fig:gCCE1 var Delta}
\end{center}
\end{figure}

\begin{figure}
\begin{center}
    \includegraphics[width=0.4944\textwidth]{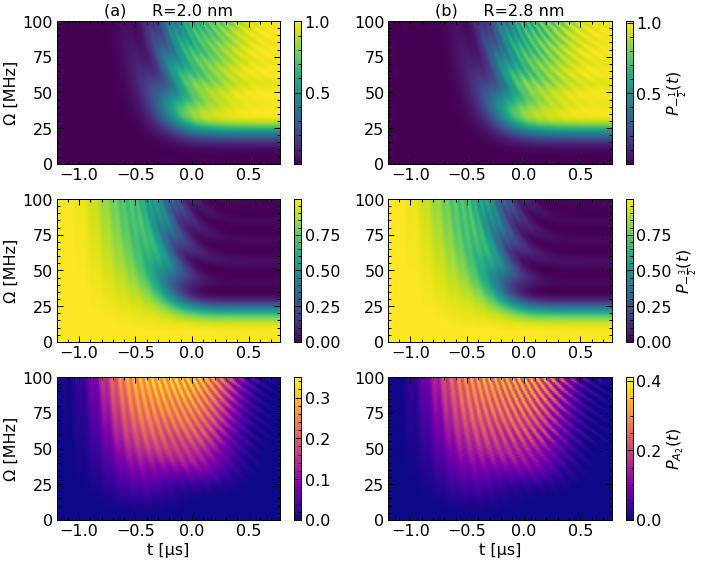}
    \caption{Filled contour plot showing the gCCE1 populations of the electron spin's $\vert\mathrm{g},-1/2\rangle$, $\vert\mathrm{g},-3/2\rangle$ and $\vert A_2\rangle$ states as a function of time and laser amplitude $\Omega$. The bath is spherical with a radius $R=2.0$ nm (a) or $R=2.8$ nm (b). In all cases $B=10$ mT and $\Delta=50$ MHz.}
    \label{fig:gCCE1 var Omega}
\end{center}
\end{figure}

It is important to underline that what we have found is only \textit{one} of the possible sets of parameters for a perfect STIRAP. The others can be found by varying the free parameters at our disposal, $B$, $\Delta$ and $\Omega$. E.g., in Fig. \ref{fig:gCCE1 var B} we show the populations as a function of both time and magnetic field in a filled contour plot, for two different values of the bath radius. Notice that the STIRAP process does not depend on the magnetic field, which is witnessed by the populations remaining the same along the y axis. This is expected behavior in that a change in the magnetic field has no effect on the dynamics, since, by examining Eq. \ref{eq:final_ham}, we discover that the energy levels associated to the $\vert\mathrm{g},-1/2\rangle$, $\vert\mathrm{g},-3/2\rangle$ and $\vert A_2\rangle$ states have no dependence on $B$. Furthermore, also the pump and Stokes pulses are not altered in the process. Note that also the difference in populations between the two bath radii is qualitatively absent. By comparing Figs. \ref{fig:gCCE1 var B}a and \ref{fig:gCCE1 var B}b we see that for $R=2.8$ nm the wiggles in $P_{-1/2}$, $P_{-3/2}$ and $P_{A_2}$ are slightly more pronounced but otherwise unchanged. Therefore, a small/medium sized bath for which the semiclassical approximation for Eqs. \ref{eq:dip_hyp_tens} and \ref{eq:dip_tens} holds does not disturb the standard execution of the STIRAP process. For both radii and for $B=10$ mT the transfer of population is perfect, which is what we wanted to achieve.

Another parameter we can control and whose variation yields important insight into the efficiency of STIRAP is the detuning $\Delta$. Regarding this, we have plotted the electron spin's populations as a function of time and detuning as filled contours in Fig. \ref{fig:gCCE1 var Delta}, for two different bath radii. Notice once again the common behavior constituted by the wiggles in the populations, which this time depend upon the chosen value of $\Delta$. For instance, the wiggles progressively disappear for $50\leq\Delta<75$, which is a sweet spot region for $\Omega=35.6$ MHz. Contrarily, whenever $\Delta$ is much smaller than the laser pulses' amplitude $\Omega$, e.g. $\Delta\sim1$ MHz, the $\vert A_2\rangle$ state is almost degenerate with $\vert\mathrm{g},-1/2\rangle$ and $\vert\mathrm{g},-3/2\rangle$. As a consequence, $\vert A_2\rangle$ is periodically excited during the application of STIRAP, thus undermining the efficiency of the population transfer. On the other hand, if $\Delta\geq75$ MHz, the laser amplitude we have chosen is not enough to ensure a perfect exchange of population and $P_{-3/2}$ does not tend to zero. Concerning the effect of the nuclear spin bath, the one containing $\sim300$ nuclear spins ($R=2.8$ nm) shows once again the absence of a qualitative change in the populations, thus corresponding to a perfect STIRAP as well, by keeping the other parameters unchanged.

Finally, the laser pulses' amplitude is an additional parameter we can vary in order to shed light on the system plus bath STIRAP dynamics. The electron spin's populations as a function of time and laser amplitude $\Omega$, for two different nuclear spin bath dimensions, can be found in Fig. \ref{fig:gCCE1 var Omega}. Note that for a detuning $\Delta=50$ MHz the population inversion due to the pulses happens only beyond a certain threshold of $\Omega_{\mathrm{threshold}}\approx\Delta/2=25$ MHz. For $\Omega<\Omega_{\mathrm{threshold}}$ the amplitude is not enough to satisfy the hypothesis of the adiabatic theorem, whereas for $\Omega\geq\Omega_{\mathrm{threshold}}$ the exchange is correctly carried out. However, in the $\Omega\geq\Omega_{\mathrm{threshold}}$ region strips appear where the efficiency of STIRAP decreases, the first one being located at $\Omega\approx40$ MHz. The other ones fall from $40$ to $85$ MHz at discrete regular steps of $15$ MHz. Furthermore, at each new step the time instant in which the transfer occurs gets slightly shifted to the right. That is why the most convenient value of $\Omega$ that has to be chosen to achieve a perfect STIRAP lies between $25$ and $40$ MHz, which contains our pick, $\Omega=35.6$ MHz, coinciding with the value obtained via the procedure in Ref.~\cite{dogra2022perfect} (Subsec. \ref{subsec:methods}). Similarly to other cases, the common behavior is constituted by the wiggles in the electron spin's populations, which are caused by the pump and laser pulses' frequencies and are not appreciably modified by enlarging the nuclear spin bath. Therefore, this type of bath has no practical effect on the STIRAP efficiency, thereby freeing experiments similar to the one described in Ref.~\cite{yale2016optical} from the decision of using NV centers in \textit{purified} diamond. Moreover, diamond has the advantage of containing only one species of paramagnetic nuclear spin, i.e. $^{13}$C. Our work is nonetheless concerned with a $\mathrm{V}_{\mathrm{Si}}^{-1}$ in 4H-SiC because of other advantages it has over diamond, related to engineering applications~\cite{mohapatra2021comparative} or technology in general~\cite{castelletto2022silicon}. It is the case to underline that the efficiency of the initialization, encoded in the infidelity $\epsilon$, can be arbitrarily chosen and the consequent set of parameters for perfect STIRAP can be found in a similar manner. We perform a population transfer with a theoretical fidelity of $99.99$\%. However, it is important to emphasize that our analysis does not include features of the system such as linewidth, power broadening, the role of meta stable states etc.

\begin{figure*}
\begin{center}
    \includegraphics[width=1.0\textwidth]{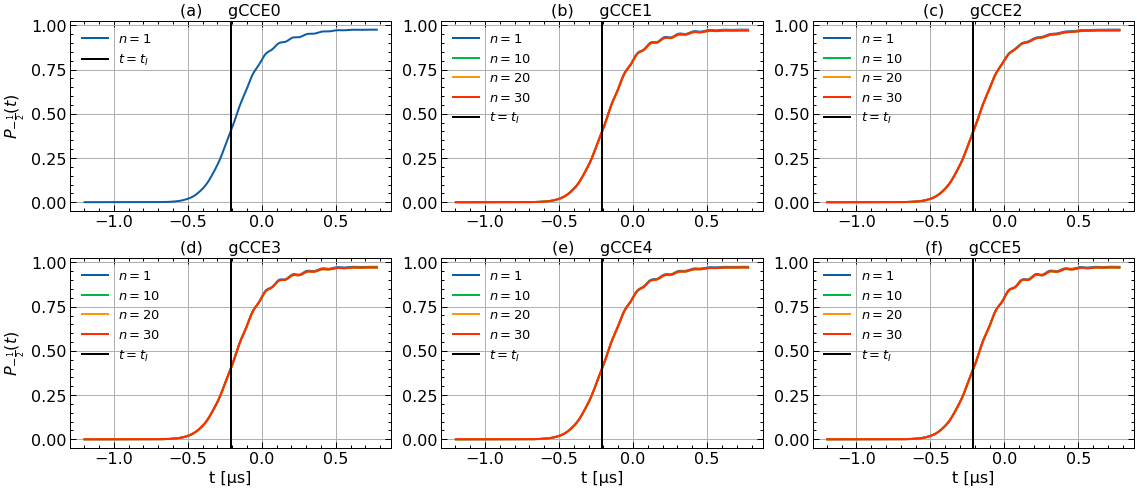}
    \caption{Population of the electron spin's $\vert\mathrm{g},-1/2\rangle$ state as a function of time, for different orders of gCCE. $P_{-1/2}$ is averaged over different numbers $n$ of realizations of the random bath-generating procedure. The bath is spherical with a radius $R=2$ nm ($\sim100$ nuclear spins) and the external magnetic field is $B=10$ mT. The other parameters are chosen from the set giving a perfect STIRAP.}
    \label{fig:gCCE_var_n}
\end{center}
\end{figure*}

\section{\label{sec:discuss}Discussion}

\begin{figure*}
\begin{center}
    \includegraphics[width=1.0\textwidth]{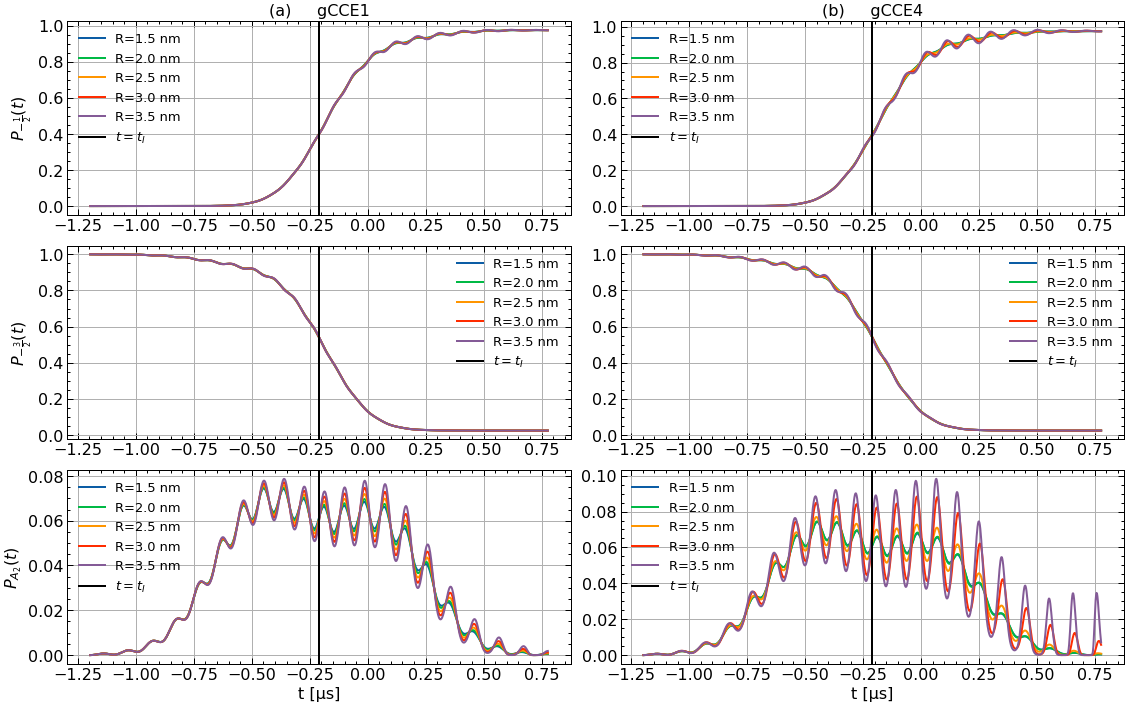}
    \caption{Populations of the electron spin's $\vert\mathrm{g},-1/2\rangle$, $\vert\mathrm{g},-3/2\rangle$ and $\vert A_2\rangle$ states as a function of time, for gCCE1 and gCCE4. The bath is spherical and the external magnetic field is $B=10$ mT. The other parameters are chosen from the set giving a perfect STIRAP.}
    \label{fig:gCCE1_4_var_R}
\end{center}
\end{figure*}

We have presented a scheme for STIRAP applied on a system interacting with a spin bath, and the gCCE method as an appropriate technique to study this kind of situation. Then we have specialized our analysis to the level structure associated to a defect formed by a negatively charged silicon vacancy ($\mathrm{V}_{\mathrm{Si}}^{-1}$) in 4H-SiC in order to prepare the associated electron spin in the $\vert\mathrm{g},-1/2\rangle$ state after initialization in the ground state, $\vert\mathrm{g},-3/2\rangle$. This task could be important for quantum control applications. First, we have considered the Hamiltonian describing the defect plus nuclear spin bath system and the drive via the pump and Stokes laser pulses, and performed  the rotating wave approximation. By analyzing this Hamiltonian, given in Eq. \ref{eq:final_ham}, we have concluded that, if the ground state is initially prepared and the already mentioned laser pulses are applied, the $\vert\mathrm{g},1/2\rangle$ and $\vert\mathrm{g},3/2\rangle$ states cannot be excited during the dynamics. As a consequence, our central five-level spin system effectively behaves as a three-level system. Then, in order to probe the bath's effect on the central spin, also considering possible intra-bath correlations, we have utilized the generalized CCE theory to divide the total effect in cluster-correlation contributions of differing size~\cite{yang2020longitudinal} (details in Subsec. \ref{subsec:methods}). By calculating the defect's populations from gCCE1 to gCCE5, never obtaining bath effects already at gCCE1, we have demonstrated that, in the range of parameter values we have chosen, the dynamics is only affected by the laser part of the Hamiltonian and the bath does not alter the process. Furthermore, we have found a set of parameter values for which a perfect STIRAP occurs, as can be seen in Fig. \ref{fig:gCCE0 and gCCE1}. Also, a change in these parameters has a small effect on the STIRAP efficiency, by slightly undermining the population transfer and maintaining a non-negligible mixture of quantum states, so that our procedure is stable. Therefore, we have analyzed in more depth the role of the parameters by varying in turn the external magnetic field $B$, the detuning $\Delta$ and the laser pulses' amplitude $\Omega$. In particular, we have demonstrated that a change in magnetic field has no effect on the dynamics if STIRAP is to be pursued, as is clear from Eq. \ref{eq:final_ham} where none of the three effective energy levels and the drives are involved by the variation of $B$. We have shown that, for a given $\Omega$, the detuning must be chosen within a sweet spot region where it is not too small with respect to $\Omega$ to cause transfers of population towards $\vert A_2\rangle$, nor too large for the $A_2$ transition to be unreachable. We have evaluated the sweet spot region to satisfy the hypothesis of the adiabatic theorem to be $50\leq\Delta<75$ (in MHz), for $\Omega=35.6$ MHz. In all of these three cases we have encountered a common behavior by noticing wiggles in the populations. This is due to the application of pump and Stokes laser pulses and linked to their frequencies. Moreover, by increasing the radius of the spherical bath we have seen that the wiggles are not appreciably modified. Consequently, we have shown that the presence of a nuclear spin bath does not disturb the population exchange with the chosen parameter values, and thus the efficiency of the STIRAP process. This is why the decision of working with \textit{purified} diamond in Ref.~\cite{yale2016optical} does not appear a necessary condition. Finally, by including an arbitrary infidelity $\epsilon$~\cite{dogra2022perfect} in our approach we have presented a flexible initialization procedure for preparing the $\vert\mathrm{g},-1/2\rangle$ state with a fidelity of approximately $99.99$\%. Although the fidelity is theoretical, the most important aspect is the flexibility of the approach, which allows one to choose $\epsilon$. It is worth mentioning again that our specialized analysis for SiC systems takes into account the specific level structure of such systems, paving the way for more accurate investigations, but does not take into account other elements which could be significant in real experiments.

\begin{acknowledgments}
The authors gratefully acknowledge financial support from PO FSE Sicilia 2014-2020 - POC Sicilia 14-20. The authors also acknowledge financial support from the FFR grant from the University of Palermo, Italy. 
\end{acknowledgments}

\appendix

\section{\label{app:A}Convergence of bath parameters}

\begin{figure*}
\begin{center}
    \includegraphics[width=1.0\textwidth]{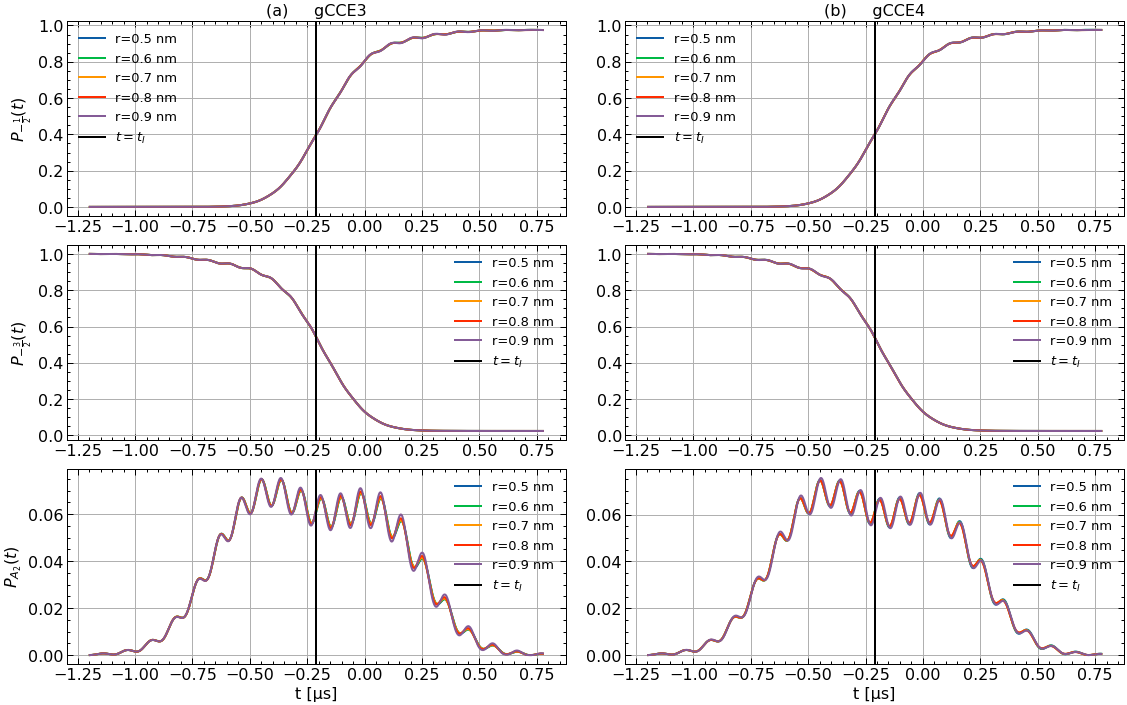}
    \caption{Populations of the electron spin's $\vert\mathrm{g},-1/2\rangle$, $\vert\mathrm{g},-3/2\rangle$ and $\vert A_2\rangle$ states as a function of time, for gCCE3 and gCCE4. The bath is spherical with a radius $R=2$ nm ($\sim100$ nuclear spins) and the external magnetic field is $B=10$ mT. The other parameters are chosen from the set giving a perfect STIRAP.}
    \label{fig:gCCE3_4_var_r}
\end{center}
\end{figure*}

In this appendix we discuss the convergence properties of the bath numerical parameters, which are the bath radius $R$, the nuclear spin connectivity $r$ and the number of bath realizations $n$. By the end of the appendix we give the converged values of such parameters and highlight how changes in them affect the STIRAP process.

First of all, in Fig. \ref{fig:gCCE_var_n} we show the population of the electron spin's $\vert\mathrm{g},-1/2\rangle$ state as a function of time, for different orders of gCCE. $P_{-1/2}$ is averaged over different numbers $n$ of realizations of the random bath-generating procedure. Notice that convergence is reached already for $n=10$, which is why we have chosen this parameter value for the simulations in the main body of the paper. The other parameters are taken from the set with which one obtains a perfect STIRAP, introduced in Sec. \ref{sec:application}, as can be seen by $P_{-1/2}$ approaching one for $t\rightarrow t_f$.

At this point, another important parameter whose convergence study yields useful information regarding the bath is the bath radius $R$. In particular, in Fig. \ref{fig:gCCE1_4_var_R} we give the populations of the electron spin's $\vert\mathrm{g},-1/2\rangle$, $\vert\mathrm{g},-3/2\rangle$ and $\vert A_2\rangle$ states as a function of time, for gCCE1 and gCCE4. As already stated in the main text, a change in the radius $R$ has a small impact on the populations for a small/medium sized bath, but for larger baths ($R>2.8$ nm) the effect is instead visible. The effect consists in the enlargement of the wiggles in the populations, and the value of the radius we have opted for is $R=2$ nm, which we have utilized throughout the numerical part of our work.

Finally, the last bath parameter we have varied is the nuclear spin connectivity $r$. In Fig. \ref{fig:gCCE3_4_var_r} we present the populations of the electron spin's $\vert\mathrm{g},-1/2\rangle$, $\vert\mathrm{g},-3/2\rangle$ and $\vert A_2\rangle$ states as a function of time, at the gCCE3 and gCCE4 orders of approximation. The variation of connectivity has little effect on the populations, which is further proof of the fact that, with the parameter values we have chosen, the presence of this type of bath has no effect on the STIRAP process. The value of the connectivity used in our work is $r=0.8$ nm, as in Ref.~\cite{seo2016quantum}.

\newpage

\bibliographystyle{unsrt}
\bibliography{apssamp.bib}

\end{document}